\documentclass[letter,12pt]{article}
\usepackage{lineno}
\usepackage{amsmath}
\usepackage{graphicx}
\usepackage{fullpage}
\usepackage{appendix}
\usepackage{natbib}

\title{Measuring global mean sea level changes with surface drifting buoys}
\author{Shane Elipot \\ Rosenstiel School of Marine and Atmospheric Science, University of Miami}

\newcommand{\Cov}{\text{Cov}}
\date{}

\begin{document}
\maketitle

\begin{abstract}
Combining ocean model data and in-situ Lagrangian data, I show that an array of surface drifting buoys tracked by a Global Navigation Satellite System (GNSS), such as the Global Drifter Program, could provide estimates of global mean sea level (GMSL) and its changes, including linear decadal trends. For a sustained array of 1250 globally distributed buoys with a standardized design, I demonstrate that GMSL decadal linear trend estimates with an uncertainty less than {0.3 mm yr$^{-1}$} could be achieved with GNSS daily random error of 1.6 m or less in the vertical direction. This demonstration assumes that controlled vertical position measurements could be acquired from drifting buoys, which is yet to be demonstrated. Development and implementation of such measurements could ultimately provide an independent and resilient observational system to infer natural and anthropogenic sea level changes, augmenting the on-going tide gauge and satellites records.
\end{abstract}

\section{Introduction}

Modern global mean sea level (GMSL) rise is an intrinsic measure of anthropogenic climate change. It is mainly the result of the thermal expansion of the warming ocean's water and the increase of ocean's mass from melting terrestrial ice \citep{Rhein2013,Church2013,frederikse2020causes}. Global mean sea level rise is a major driver of the regional \citep{Hamlington2018} and coastal sea level extremes \citep{Woodworth2015,Marcos2017} that impact millions of human lives and assets \citep{Anthoff2006,Nicholls2011}, and threatens ecosystems \citep{Craft2009}. Currently, global mean and regional sea level changes are monitored by two observational systems: coastal and island tidal gauges \citep{Douglas2001,Woodworth2003} and satellite radar altimeters \citep{Nerem1995,Ablain2017}. Tidal gauge records have high temporal resolution but their representativeness of the global mean sea level is biased towards the coasts and the Northern Hemisphere. In contrast, the altimeter reference record is almost global (covering latitudes below 66$^\circ$), but can only provide a near synoptic view about every 10 days. Here I show that if the satellite-tracked surface drifting buoys (hereafter drifters) of the Global Drifter Program recorded not only their geographical coordinates by the Global Positioning System (GPS), but also their altitudes above the mean sea surface, the drifter array could provide estimates of GMSL changes, including long-term trends. With the current size of the drifter array, provided that the impact of potential biases is apprehended and quantified, GMSL decadal linear trend estimates with an uncertainty less than {0.3 mm yr$^{-1}$} could be achieved with drifter GPS altitude random errors better than 1.6 m. The drifter array could thus provide an independent and resilient observational system to infer natural and anthropogenic sea level changes, not only validating and augmenting the on-going tide gauge and satellites records, but also introducing an important redundancy that would preserve the continuity of the record in case of failure of key satellite missions.

\section{Concept and Methods}

In order to demonstrate the feasibility of the proposed idea, I take the example of NOAA's Global Drifter Program. Previously relying uniquely on the Argos system \citep{lumpkin2007measuring}, this drifter array has almost completely transitioned to the Global Positioning System (GPS) for localization purpose, and to the Iridium satellite system to transmit their data to data centers on land \citep{elipot2016global}. Yet, despite the altitude coordinate being part of the positioning solution, only latitude and longitude data are retained in the transmitted data stream. 

GPS altitude data at sea are a measure of height of the sea surface above a reference ellipsoid which center coincides with the Earth's center of mass. The relevant quantity is rather the height of the sea surface relative to a reference mean sea surface (MSS). The MSS is the sum of an equigeopotential surface called the geoid and a mean dynamic topography associated with time-mean balanced oceanic motions over a reference time period. The interest here lies in relatively low-frequency changes of the mean sea level (MSL) which is the sea surface height (SSH) averaged over a given time period such as a day or a month. MSL is of particular interest when further averaged in space, regionally or globally, as its changes integrate physical processes associated with ocean dynamics and thermodynamics, such as its volume changes and its overall rise as a result of global climate change. As a consequence, GPS data acquired by a drifter would need to be subtracted from a co-located estimate of a reference time-invariant MSS. The primary signal of interest is the local MSL
\begin{linenomath*}\begin{equation}
\label{eq1}
  h_{MSL} = h_{GPS} - h_{MSS} + \varepsilon,
\end{equation}\end{linenomath*}
where $\varepsilon$ is an error, the sum of all physical contributions to SSH variability that are not relevant for global or regional MSL changes but may still affect their estimates. Such contributions include surface gravity waves, astronomically-forced tides, and internal gravity waves. The variance over time scales shorter than the time scale defining the MSL, in other words the uncertainty of $h_{MSL}$ that would be experienced by a drifter, is
\begin{linenomath*}\begin{equation}
\label{eq2}
  \text{Var}(h_{MSL}) = \text{Var}(h_{GPS}) + \text{Var}(h_{MSS}) + \text{Var}(\varepsilon) + 2\Cov(h_{GPS},\varepsilon),
\end{equation}\end{linenomath*}
since the MSS is independent of {the hypothetical} GPS heights from drifters. The last term of the above expression is the covariance between GPS height measurements and sea level variability which is not relevant for MSL. The purpose of this study is to simulate $h_{MSL}$ and to quantify {the individual contributors to} $\text{Var}(h_{MSL})$ for drifters, and subsequently {to} calculate their global averages in order to assess if the drifter array could monitor GMSL changes with sufficient accuracy. {The Global Climate Observing System (GCOS)  recommends in its 2016 implementation plan} \citep{GCOS} {that GMSL and GMSL decadal trend, two Essential Climate Variables, be reported on weekly to monthly time scales with an absolute stability, or uncertainty, of 2-4 mm and $0.3$~mm~yr$^{-1}$, respectively. In this paper, I assess if such levels of accuracy could be achieved with the drifter array. As a point of comparison, one comprehensive and recent assessment reveals that GMSL decadal linear trend uncertainties from altimeter data vary between 1 and 0.5 mm~yr$^{-1}$} [Figure 5 of \citet{ablain2019uncertainty}].

In order to simulate $h_{MSL}$, I construct and analyze synthetic MSL measurements from hourly drifter trajectories \citep{elipot2016global} using sea surface height data from the GLORYS2V4 ocean reanalysis (Appendix A and Figure A1), excluding data in waters shallower than 120 m and poleward of 66$^\circ$ N and S. This ocean model assimilates sea level anomalies (SLA) from altimetry and in situ profiles of temperature and salinity, and thus represents MSL changes due to both steric and non-steric (mass) changes \citep{gregory2019concepts}, yet imperfectly which is a caveat of this study \citep{Ferry2012,desporte2017quality,garric2018performance}. Since ocean models implicitly assume that the geoid and the reference ellipsoid coincide \citep{gregory2019concepts} the time mean of the {GLORYS2V4} sea surface height is the mean dynamic topography portion only of the MSS, and therefore vertical land motions and geoid changes are not modeled \citep{tamisiea2011moving}. As a result, using SLA calculated as deviations from the time mean of the modeled sea surface height simulates $h_{GPS} - h_{MSS}$ [eq.~(\ref{eq1})] in a geocentric reference frame. This implies the need to ultimately convert global-mean geocentric sea level measured by the proposed drifter array (or altimetry) to global-mean sea level rise (or changes) by accounting for the effect of contemporary deformations of the Earth's surface, including global isostatic adjustment (GIA), and redistribution of mass \citep{frederikse2017ocean,gregory2019concepts}. This conversion takes the form of an added component to the GMSL trends with an associated uncertainty which is ultimately added to my final error budget for GMSL decadal linear trend \citep{frederikse2019imprints}. But as a first step, I linearly interpolate in time and space daily SLA from the GLORYS2V4 ocean reanalysis along drifter trajectories, and I subsequently calculate a global mean at daily time steps to produce a relative time series of drifter GMSL changes [Appendix B, \citep{Henry2014,Masters2012}]. Does such a drifter GMSL time series represent true GMSL changes? To answer this question, this time series is compared next to a reference GMSL time series which is calculated from the ocean reanalysis using all grid points passing the same spatial selection criteria as the drifters.

\section{Error analysis and budget}

\subsection{Sampling error ($\sigma_{\text{S}}$)}
\label{sec:sampl-error-sigm}

The rise of GMSL induced by anthropogenic forcing is of primary importance, yet its estimate is still uncertain \citep{nerem2018climate,kleinherenbrink2019revised}. Can the drifter array capture the GMSL upward trend with sufficient accuracy despite its heterogeneous sampling? After subtracting seasonal cycle estimates (Appendix B), using daily time series, the linear trend of the drifter GMSL estimate based on model data is $2.85 \pm 0.12$~mm~yr$^{-1}$ between 1993 and 2015. In this paper, reported uncertainties correspond to standard errors (1-$\sigma$) or 68\% confidence intervals, and trends are estimated taking into account serially correlated noise [Appendix B, \citet{bos2013fast}]. For the same period, the reference GMSL trend from the model is $3.30 \pm 0.02$~mm~yr$^{-1}$ which is equal to an altimetry-based estimate [see Figure 9 of \citet{ablain2019uncertainty}]. The drifter GMSL therefore underestimates the true linear trend over 23 years by $0.45$~mm~yr$^{-1}$, but this is because of a large positive bias between 1993 and 1995 when the number of drifters is small (Figure \ref{fig1}a,c). Therefore, for the remainder of this paper I consider statistics and linear trends over the last decade of simulated data only, from the beginning of 2006 to the end of 2015. The year 2006 corresponds to the time when the drifter array reached maturity \citep{lumpkin2017advances} with approximately 1250 drifters (Figure \ref{fig1}c).

For the period 2006-2015, the rms difference between the GMSL daily time series, after subtracting seasonal cycle estimates, is 4.4 mm, which is counted independently for the total error budget for GMSL. This error, not formally represented in eq. (\ref{eq2}) because it arises from the global averaging calculation, is the estimated sampling error, or the inability of the drifter array to capture the true global average of the MSL as a function of time because of the spatial inhomogeneity of the array. The drifter GMSL trend for 2006-2015 is $3.87 \pm 0.15$~mm~yr$^{-1}$, not significantly different from the model reference trend of $3.81 \pm 0.1$~mm~yr$^{-1}$ [or the altimeter-based trend, \citet{ablain2019uncertainty}]. The higher trend values estimated from the later part of the record are a consequence of the acceleration of GMSL rise \citep{nerem2018climate,ablain2019uncertainty}, effectively rendered by the ocean reanalysis. Yet, the trend estimate of the difference daily time series (model minus drifters, Figure \ref{fig3}a) is $-0.05 \pm 0.08$~mm~yr$^{-1}$, which represents a bias and random error which leads to a total trend error of $0.1$~mm~yr$^{-1}$ from the sampling error.

\subsection{GPS errors ($\sigma_{\text{GPS}}$)}
\label{sec:gps-errors-sigm}

The technical limitation of this study is that it is assumed that altitude measurements with sufficient accuracy can be acquired regularly from relatively small buoys drifting at sea [the GPS receivers currently in use have an overall estimated horizontal accuracy of 22 m \citep{elipot2016global}]. This requires that enough computing power is present onboard each drifter to regularly calculate and transmit an accurate 3D position solution, or that  time series of high frequency position data (typically at 1 Hz) could be transmitted intermittently to land for post-processing. Either solutions requires more electric power than currently available on a drifter to retain its 450-day designed life expectancy \citep{lumpkin2016fulfilling}. In the rest of this study, I assume that this power issue is solved, which is not unreasonable considering our current state of technological innovation. This study is therefore an ocean observing simulation, and is worthwhile as such, especially considering the critical importance of further measuring and understanding regional and global MSL changes. 

It is necessary to consider the systemic errors from GPS contributing to the terms $\text{Var}(h_{GPS})$ and $2\,\Cov(h_{GPS},\varepsilon)$ in eq. (\ref{eq2}). These errors include but are not limited to issues such as atmospheric effects, satellite orbit mismodeling, satellite and receiver clocks precision, the choice of GPS calculation method \citep{santamariagomez2011correlated}, and the unkown effects of the sea state on GPS height measurements from drifters. Despite such issues, GPS-equipped buoys at sea are successfully used to calibrate satellite radar altimeters, providing estimates of absolute biases of altimeter SSH data with an accuracy now reaching between 1 and 2 cm \citep{Frappart2015,Born1994,Key1998,Watson2003}. Historically, SSH measurements by GPS have relied on differential GPS techniques for which the GPS height of a buoy is calculated relative to a reference GPS station on land, typically no more than a few kilometers away \citep{Watson2003,Watson2011}. Such differential GPS method could not therefore be applied to the global drifter array for the purpose of estimating GMSL. Recently, new GPS methods relying on a single GPS receiver platforms have provided the opportunity to forgo reference land stations and therefore obtain SSH measurements in the absolute geocentric reference frame over the open ocean \citep{Fund2013,Chen2013}. In particular, the Precise Point Positioning (PPP) method can provide SSH measurements with accuracy as high as 5 cm when compared to 6-min tide gauge measurements \citep{Kuo2012}. In fact, the PPP method has been successfully applied as an example to measure the surface height gradient of the Loch Ness with an unmanned water surface vehicle  \citep{moralesmaqueda2016water}, or to calibrate altimeter data with accuracy better than 2 cm \citep{Frappart2015}. Whereas dedicated calibration experiments can rely on high frequency GPS data (typically at 1 Hz) and ample resources for post-deployment processing, the instrumental setup of drifters for accurate GPS measurements will likely be constrained by issues such as of battery capacity, on-board processing power and methodologies, and available bandwidth for transmitting position data. As a consequence of the lack of existence of actual data, I do not attempt here to assign unique values to the GPS error terms, but rather I consider a range of values for the overall GPS errors per individual daily drifter measurement, up to 2 m standard deviation for daily averages---a reasonable upper bound---and study the impact of all combined errors as a function of the number of drifters in the array. 

\subsection{Geophysical random errors ($\sigma_{\text{SSH}}$)} 
\label{sec:geophys-rand-errors}

Variance in daily MSL from hourly heights acquired by drifters will arise from relevant physical processes not represented by the model of the GLORYS2V4 ocean reanalysis: SSH variability from barotropic tides, internal waves including baroclinic tides, and surface gravity waves.

\subsubsection{Tides and internal waves} SSH variability from barotropic and baroclinic tides and internal gravity waves is estimated here using maps of SSH variance from a global 1-year run of the HYCOM numerical model forced by atmospheric fields and tidal potential \citep{Savage2017}. Specifically, SSH variances from the diurnal, semi-diurnal and supertidal frequency bands are interpolated at all drifter hourly locations (Figure A2, median value of corresponding standard deviations is 0.19 m), and subsequently globally-averaged to derive a daily time series of error (Figure \ref{fig1}b). From 1993 to 2006, this error decreases steadily as the number of drifters increases. After 2006, the error time series is strongly anticorrelated with the number of drifters (at -0.90), averaging to 1.6 mm. When compared to a discrete number of in-situ observations, HYCOM can underestimate the observed high-frequency SSH variance \citep{Savage2017}, thus the MSL errors from these processes may be underestimated here. However, these estimates may still be adequate since a fraction of the stationary component of the SSH variability in tidal bands could be subtracted from drifter observations using predictions from a  barotropic tidal model \citep{EgbertE02}, and the error used here could be actually reduced. In addition, on-going research on the stationary component of baroclinic waves suggest that part of their SSH signal may also be subtracted \citep{Zaron2017,zaron2019baroclinic}.

\subsubsection{Surface Gravity Waves} SSH variability from surface gravity waves is estimated using significant wave height (SWH) data obtained from a global, 0.5$^\circ$ spatial resolution, 3-hr temporal resolution, run of the wave model WAVEWATCH~III as part of the IOWAGA project (Appendix A), forced by National Centers for Environmental Prediction Climate Forecast System Reanalysis winds \citep{Ardhuin2011}. The SSH variance from surface gravity waves calculated as $\text{(SWH)}^2/16$ is linearly interpolated at hourly time steps along the drifter trajectories (Figure A2, median value of corresponding standard deviations is 0.58 m), and subsequently globally-averaged to derive a daily time series of error (Figure \ref{fig1}b). From 1993 to 2006, this error also decreases as the number of drifters increases. After 2006, the error time series is anticorrelated with the number of drifters (at -0.71), averaging to 4.4 mm.

\subsubsection{Mean Sea Surface error} 
\begin{figure}
\centering
  \noindent\includegraphics[height=0.8\textheight]{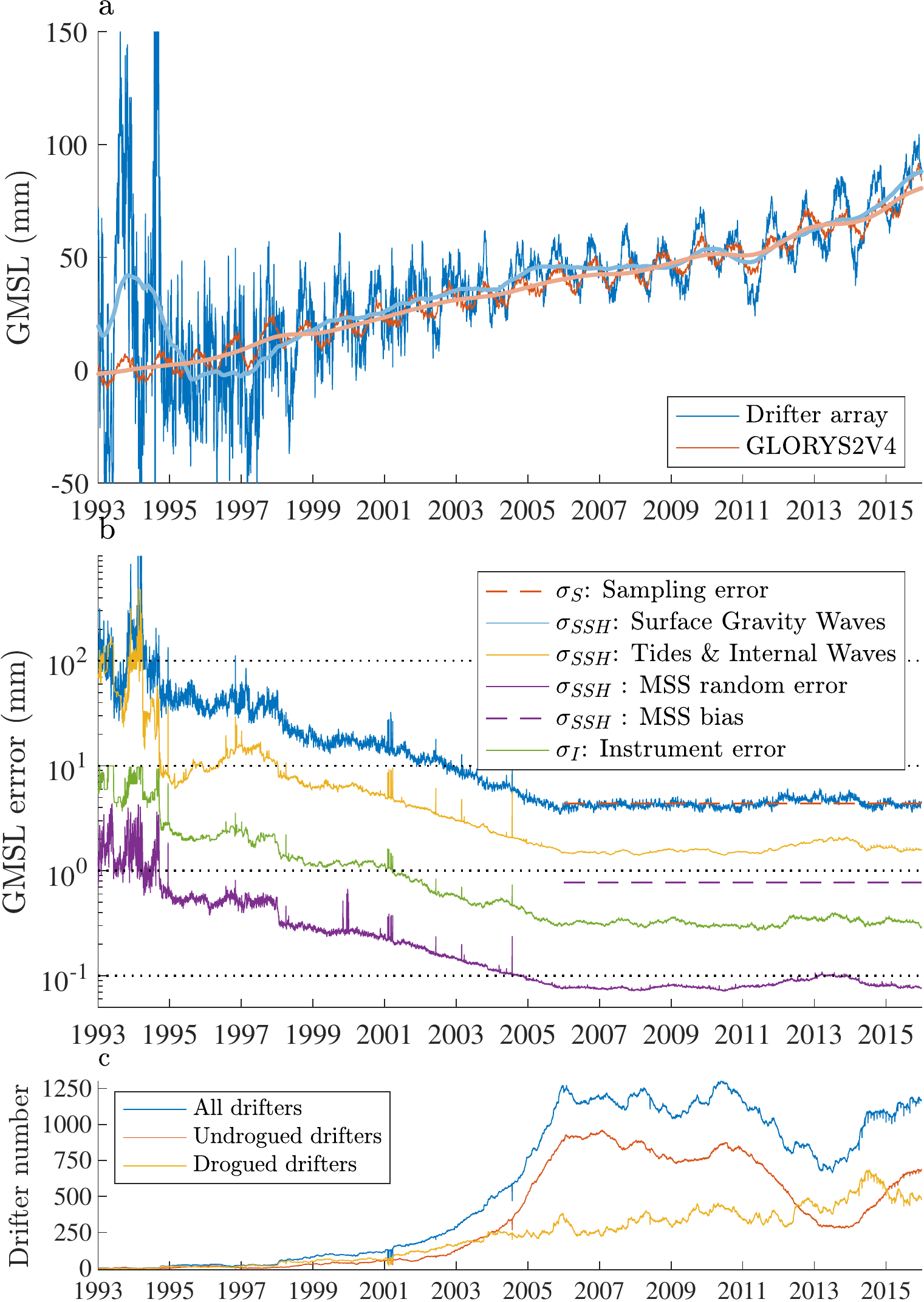}
  \caption{\textbf{Global mean sea level estimates}. a: GMSL time series from simulated drifter data and GLORYS2V4 ocean reanalysis data excluding data in water depth less than 120 m and poleward of 66$^\circ$ north and south. The thicker lines are the 9-month smooth estimates after removal of seasonal cycles. b: Daily error estimates for the drifter GMSL time series as defined in the text. c: Drogued, undrogued, and total number of drifters per day in the hourly drifter database version 1.01.}  
  \label{fig1}
\end{figure}

SSH errors will also arise from uncertainties in the MSS subtraction (eqs. \ref{eq1} and \ref{eq2}), and are quantified by considering that these errors have two contributions: representation errors that include MSS modeling or methodological biases \citep{Pujol2018}, and random errors arising from the errors in the data used for mapping. To quantify representation errors, I consider two MSS products, CNES CLS15  \citep{Schaeffer2012} and DTU15 \citep{Andersen2015}, which use overlapping altimetry datasets but different methodologies. I linearly interpolate the time-invariant but spatially-varying difference of MSS (CLS15 minus DTU15) along drifter trajectories (Figure A2, 8.65 mm median absolute value), and calculate a global average at daily time steps to produce a time series of MSS difference (Figure \ref{fig2}b). The rms value of this time series, 0.77 mm for 2006-2015, is interpreted as a representation error, or bias, for the MSS (Figure \ref{fig1}b).  To quantify MSS random errors, I use the time-invariant formal error map of the CNES CLS15 MSS product that I also interpolate at drifter positions (Figure A2, 11 mm median value), and subsequently globally-average to derive a daily time series of error (Figure \ref{fig1}b). After 2006 this random error time series is anticorrelated (at -0.96) with the number of drifters and averages to 0.08 mm.

\subsection{Instrument bias ($\sigma_{\text{I}}$)} 
\label{sec:instr-bias-sigm}

The next systematic source of errors to consider is an instrument-to-instrument bias due to the positioning of the GPS antenna within the buoy of an individual drifter. To measure sea level, the use of a buoyant float containing a GPS antenna requires to evaluate the vertical distance between the flotation line and the GPS antenna phase center. Using reference tide gauge data, calibration of GPS antenna on buoys can provide such distance with an accuracy better than 1 cm \citep{Frappart2015}. However, in the case of the drifter array, calibration of all buoys would not be practically feasible so that an uncertainty will arise from unavoidable deviations in the construction of each drifter, implying a constant bias for the life of a single drifter. I assess the potential impact of such instrument-to-instrument bias by assigning to each individual drifter a random constant bias drawn from a normal distribution with a standard deviation of 1 cm, and subsequently recalculate the drifter GMSL. This operation is conducted one hundred times and the variance of all these alternate GMSL time series is calculated at daily time steps (Figure \ref{fig2}c). This variance takes the form of an added random error, averaging 0.32 mm for 2006-2015. This variance time series is anticorrelated (at -0.67) with the number of drifters.

\subsection{Surface Wave bias} 
\label{sec:surface-wave-bias}

Another possible source of errors arises from the fact that the surface float of a drogued drifter has been observed to sink underwater as wave crests passe its location, being pulled underwater by the tethered drogue [R. Lumpkin, personal communication]. As the height of a drifter is eventually calculated by the GPS from data acquired over a finite time-window every hour (the exact processing varies between the different drifter manufacturers), the sinking behavior of the buoy could introduce a negative bias as a drogued drifter would sample preferentially the troughs over the crests of surface gravity waves. I quantify this potential bias by considering again the outputs of a numerical simulation of waves \citep{Ardhuin2011}, but this time taking the standard deviations of the wave field as being the scale parameters for probability distribution functions which are originally normal and centered---a reasonable first approximation for SLA induced by waves \citep{longuethiggins1963}---but with all values above one standard deviation cut off. The new means of such truncated distributions effectively provide the biases which are linearly interpolated in time and space at drogued drifter positions only. The wave bias is ignored for undrogued drifters as these ones are expected to ride the waves. The true in situ behaviors of drifters are actually unknown, but the two extreme behaviors considered here should be representative of the range of drifter behaviors. The impact of these biases on trend estimates suggest that in the presence of wave biases, it will be necessary to calculate separately GMSL estimates for drogued and undrogued drifters. Indeed, averaging simulated data from both biased drogued drifters and unbiased undrogued drifters leads to an overall GMSL fall, rather than a rise, during for the 2006-2015 period (Figure \ref{fig2}b, yellow curve). This occurs because during approximately the years 2013 and 2014 the number of drogued drifters surpassed the number of undrogued drifters (Figure \ref{fig1}c). In contrast, using unbiased undrogued drifters only (on average 670 daily over the last decade) provides a GMSL decadal linear trend of $3.72 \pm 0.12$~mm~yr$^{-1}$, an insignificant error of only -0.09~mm~yr$^{-1}$ compared to the model reference trend. Using biased drogued drifters only (on average 394 daily) provides a GMSL decadal linear trend of $4.24 \pm 0.21$~mm~yr$^{-1}$, an error of 0.43~mm~yr$^{-1}$. Ultimately, the impact the wave bias becomes an issue of sampling bias and the number of drifters to be used to reliably estimate trends, considered below.

\subsection{Other errors for trend estimates}
\label{sec:other-errors-trend}

Temporal variability of the geocenter of the frame of reference for GNSS systems like GPS induces an uncertainty in long-term linear trend estimates for global-mean geocentric sea-level of 0.1 mm~yr$^{-1}$  (1-$\sigma$) \citep{couhert2015towards,santamariagomez2017uncertainty,ablain2019uncertainty}. The proposed drifter system will be susceptible to that same uncertainty and that value is adopted for this study. In addition, global-mean geocentric sea-level changes derived from satellite altimetry or from the current proposed array need to be converted to global-mean sea-level rise [the increase in the volume of the ocean divided by the ocean surface, see \citet{gregory2019concepts}]. The difference between the two metrics is due to deformations of the Earth's surface from GIA and the redistribution of mass from contemporary ice melt and hydrology, amounting for linear trend estimates to 0.27 mm~yr$^{-1}$ and 0.21 mm~yr$^{-1}$, respectively over the previous decade \citep{frederikse2017ocean,frederikse2020causes}. The estimated standard errors (1-$\sigma$) on these trends are 0.05 mm~yr$^{-1}$ and 0.02 mm~yr$^{-1}$, respectively [updated values from \citet{frederikse2019imprints}, personal communication]. The three uncertainty estimates for linear trends listed above are counted towards the total error budget for decadal linear trend.


\subsection{Error budget}

As a final step, I combine all relevant GMSL errors to derive estimates of daily and monthly GMSL uncertainties, as well as decadal trend uncertainty based on daily values, as function of the number of drifters and a range of GPS random error. The error variance of the GMSL daily estimates for the 2006-2015 time period is estimated as the sum of four terms:
\begin{linenomath*}\begin{equation}
  \label{eq:sig2}
  \sigma^2_{\text{GMSL}} = \sigma^2_{\text{S}}  + \frac{\sigma^2_{\text{GPS}}}{N} + \sigma^2_{\text{SSH}} + \sigma^2_{\text{I}},
\end{equation}\end{linenomath*}
where  $\sigma^2_{\text{S}}$ is the sampling error variance quantified in section~\ref{sec:sampl-error-sigm}, $\sigma^2_{\text{GPS}}$ is the GPS error for $N$ individual drifters on a given day discussed in section \ref{sec:gps-errors-sigm}, $\sigma^2_{\text{SSH}}$ is the sum of three contributions to the physical errors quantified in section \ref{sec:geophys-rand-errors} (tides, surface gravity waves, and MSS errors), and $\sigma^2_{\text{I}}$ is the instrumental error variance quantified in section \ref{sec:instr-bias-sigm}. Regression theory predicts that an error variance $\sigma^2$ propagates into the error variance for linear trend parameter as $ {\sigma^2}/{\sum_k(t_k - \overline{t})^2},$ where $t_k$ are the time steps for the time period over which the trend is estimated and $\overline{t}$ is the mid-point in time. Such error propagation is valid for non serially correlated errors which is reasonably assumed to be the case in the absence of real data for this study for the GPS errors, the physical errors, and the antenna errors, but not for the sampling errors as discussed in section \ref{sec:sampl-error-sigm}.

The GPS error variance for GMSL, ${\sigma^2_{\text{GPS}}}/{N}$, is a simplification of a global average calculation that neglects the area-weighting (Appendix B), and also assumes that the GPS height daily-averaged error variance $\sigma^2_{\text{GPS}}$ is the same for all drifter hourly measurements. When used to propagate into trend errors, I assume that these globally-averaged GPS errors are uncorrelated from one day to the next. GPS errors from land stations are typically correlated \citep{mao1999noise,zhang1997southern}, their spectrum being often characterized as a mixture of low-frequency white noise and high-frequency power law noise \citep{santamariagomez2011correlated}. Nevertheless, because of a lack of actual data and as a preliminary approach, I assume that daily averaging of hourly measurements of a single drifter, as well as globally averaging over all drifters, should reasonably filter out correlated errors for GMSL. This assumption will only be testable once surface drifters record and transmit their altitude data.

The SSH variance from unresolved ocean physics and MSS errors, $\sigma^2_{\text{SSH}}$, and the variance from the antenna bias $\sigma^2_{\text{I}}$, estimated at daily time steps are both positively correlated with the inverse of the number of drifters (correlations 0.78 and 0.67, respectively). Thus, to quantify the impact of the number of drifters on the proposed observing system, I derive the linear model  $\sigma^2_{\text{SSH}} + \sigma^2_{\text{I}} = \alpha_1 N^{-1} + \alpha_2$ where $\alpha_1 = 135^2$ mm$^2$ and $\alpha_2 = 2.28^2$ mm$^2$.

Finally, I use eq. (\ref{eq:sig2}) to produce a range of estimates for $\sigma_{\text{GMSL}}$ by varying the values of $N$ and  $\sigma_{\text{GPS}}$  (Figure~\ref{fig3}a). The same calculation is re-conducted for monthly GMSL values (Figure~\ref{fig3}b, Appendix B). Next, the daily GMSL errors are propagated into decadal linear trend errors by combining in quadrature the GMSL sampling error (0.1~mm~yr$^{-1}$, section~\ref{sec:sampl-error-sigm}) and the other GMSL errors. The errors due to the reference frame uncertainty and the GIA and hydrology corrections are also added in quadrature to produce the final estimates (Figure~\ref{fig3}c).

For the nominal size of the drifter array ($1250$ drifters), however small the magnitude of the GPS error is, the uncertainty for GMSL daily estimates is always larger than 6 mm and increases rapidly with the GPS error. For monthly GMSL estimates however, if the standard deviation of daily GPS error is less than 40 cm, the uncertainty remains below 4 mm which is the upper bound of the GMSL errors recommended by GCOS (2 to 4 mm). As for an estimate of the GMSL decadal linear trend based on 1250 drifters, this one  would be less than $0.3$~mm~yr$^{-1}$, the upper value recommended by GCOS, if the GPS daily error remains below 1.6 m standard deviation. If only undrogued drifters unaffected by a potential wave bias are used (670 on average for the 2006-2015), the uncertainty for the decadal linear trend would be below $0.3$~mm~yr$^{-1}$ if the GPS daily error is maintained below 1.17 m. These errors are more than an order of magnitude larger than the reported errors for geodetic buoys at sea using PPP GPS techniques \citep{Fund2013,Frappart2015}. This suggests that while GPS receivers with carefully controlled and standardized accuracy would be needed for this proposed GMSL observing system, they would not necessarily need to be of the same geodetic quality standard as the ones used for satellite altimetry calibration.

\section{Conclusions}

In conclusion, this study aimed to demonstrate that the global drifter array could provide a third means---in addition to satellite altimetry and tide gauges---to monitor GMSL changes related to climate processes, provided that drifters are tracked using carefully controlled GNSS technologies. A third observational system to measure mean sea level would be greatly beneficial to further validate and calibrate the existing observing systems, as well as refine the closing of the sea-level budget, and also provide a redundancy in case of failure of the existing systems. Importantly, drifters could provide a reliable estimate of GMSL trends related to climate change. An estimate of GMSL from drifters would not have the geographical limitations of the tide gauge network or of the reference altimeter satellites, and may provide truly synoptic GMSL estimates at higher frequency than derived from the reference altimetry satellites. The ocean observing system simulated in this study is currently not achievable because of the anticipated limitation of the GPS receivers currently equipping the drifters of the global array. This limitation cannot be fully tested because the position records of GPS-tracked surface drifters do not currently contain altitude. Considering the median lifetime of 240 days of drifters between 2006 and 2015, if the Global Drifter Program array changed its specification to equip drifters with standardized GPS receivers with a tested daily random error of 1.6 m or less in the vertical direction, and provided that potential biases are apprehended and quantified, the drifter array could be providing its first GMSL estimates within a few years, augmenting the on-going satellite and tide gauge observational networks.

\begin{figure}
  \noindent\includegraphics[width=\textwidth]{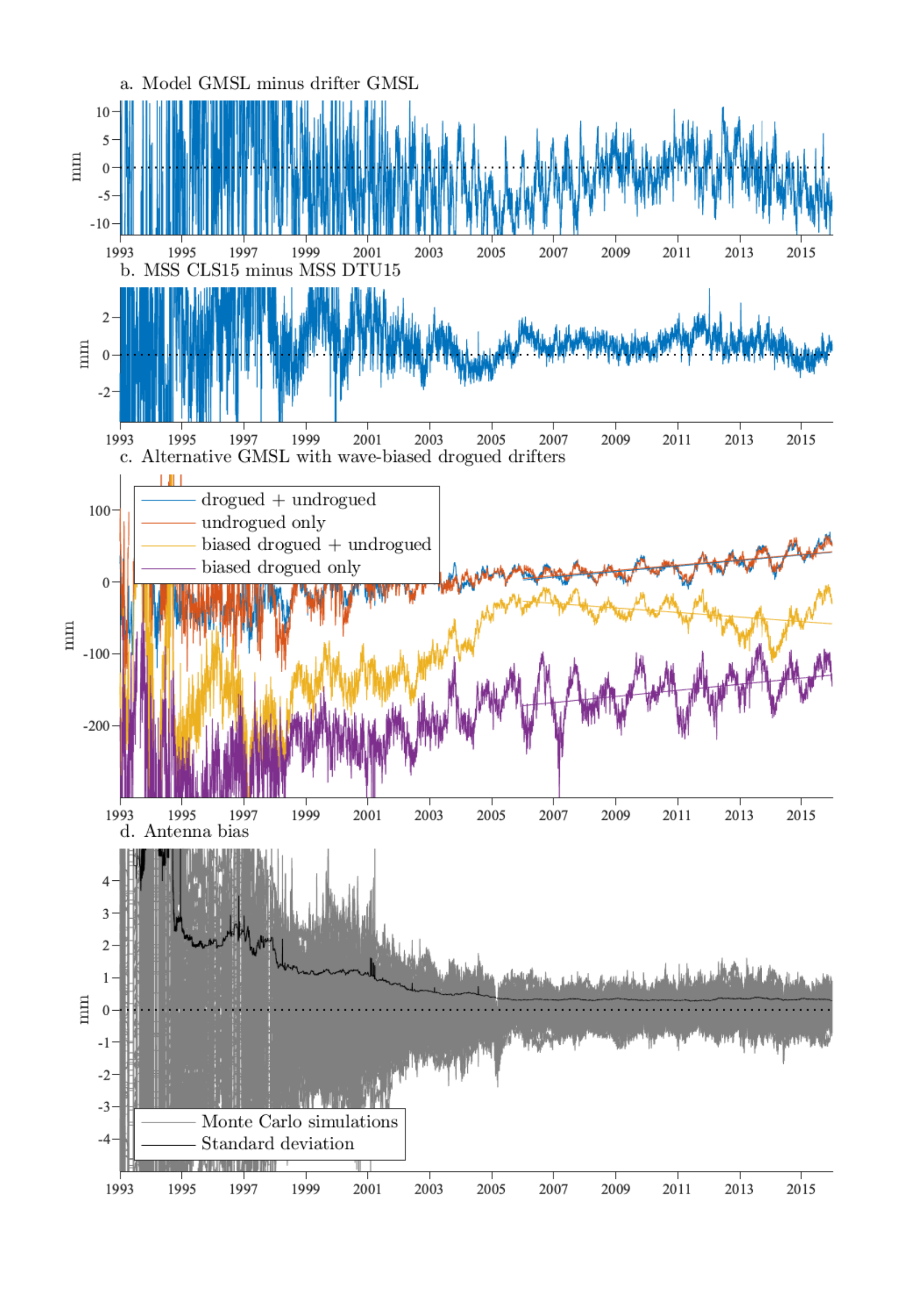}
\end{figure}

\begin{figure}
\caption{\textbf{Bias estimates}. \textbf{a. Sampling:} difference between GMSL reference time series from the model and GMSL time series from drifter simulated measurements. \textbf{b. Mean Sea Surface:} time series of globally-averaged difference between the MSS CLS15 product and the MSS DTU15 product as sampled by the drifter array. \textbf{c. Surface waves: }Various cases of GMSL estimates when using drifters with biases from surface gravity waves. The blue curve is the same drifter GMSL time series as in Figure \ref{fig1}a. The straight lines show the linear trend estimates of each curves for 2006-2015 after removing seasonal cycle estimates. \textbf{d. Antenna:} 100 recalculated GMSL times series from drifters data assuming random antenna phase center biases, minus the unbiased GMSL time series (gray curves). The black curve shows the standard deviation at daily time steps.}
  \label{fig2}
\end{figure}

\begin{figure}
  \noindent\includegraphics[width=\textwidth]{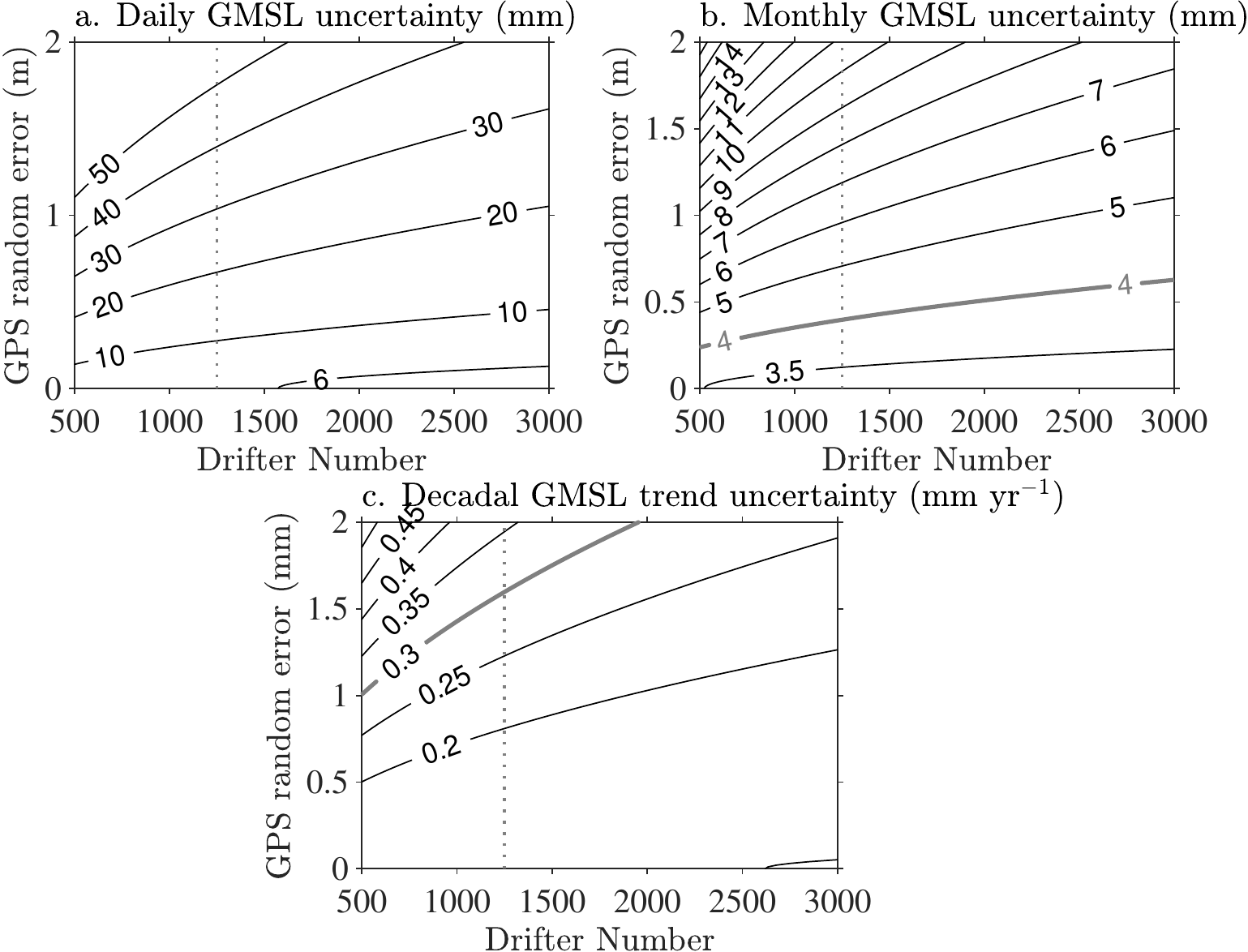}
  \caption{\textbf{Uncertainty for drifter GMSL and GMSL trend estimates}. {a.} Uncertainty in mm of daily drifter GMSL estimates as a function of the number of drifters and the individual drifter GPS altitude random error. {b.} Same as in {a.} but for monthly drifter GMSL estimates. {c.} Uncertainty in mm per year of drifter GMSL decadal linear trend based on daily estimates as a function of the number of drifters and the individual drifter GPS altitude random error. In each panel the vertical dotted line indicates 1250, the goal number for the drifter array.}
  \label{fig3}
\end{figure}

\clearpage

\appendixpage
\appendix
\setcounter{figure}{0} \renewcommand{\thefigure}{A\arabic{figure}}
\section{Data sources}
\label{A1}

The hourly drifter database \citep{elipot2016global} is available from the website of the Data Assembly Center of the Global Drifter Program, hosted at the NOAA/Atlantic Oceanographic and Meteorological Laboratory (http://www.aoml.noaa.gov/phod/dac/hourly\_data.php). Out of the database version 1.01, I selected 12,784 unique drifters between January 1, 1993 and December 31, 2015, totaling 115,344,433 hourly locations of drifters both drogued and undrogued. The hourly database contains only continuous drifter trajectory segments of 12-hr length or longer.

\begin{figure}
\centering
 \noindent\includegraphics[width=0.8\textwidth]{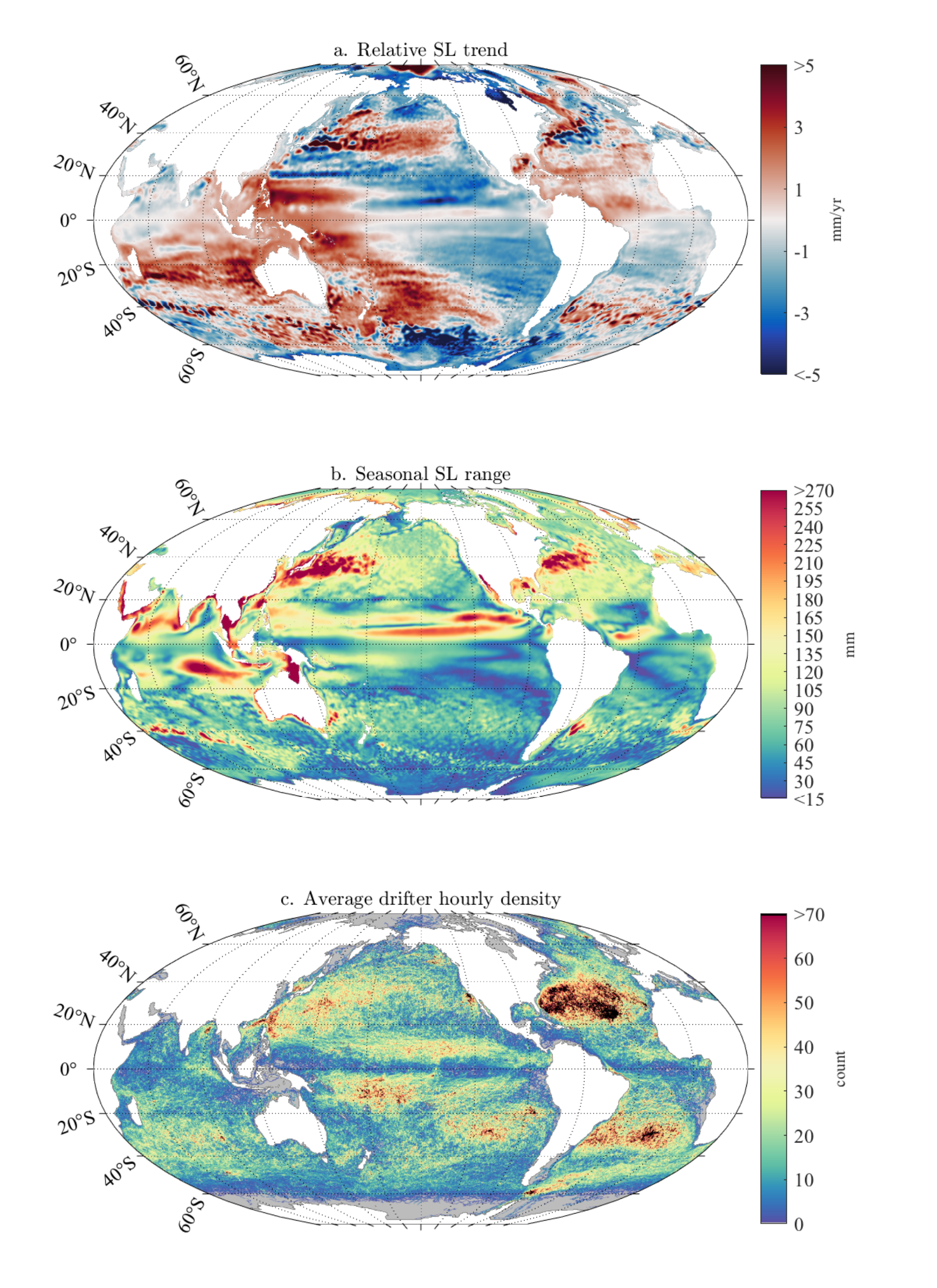}
\caption{\textbf{Mean Sea level characteristics of the GLORYS2V4 reanalysis and density of hourly drifter observations.} \textbf{a.} Local linear sea level trends from 1993 to 2015 minus the GMSL trend of 3.30 mm yr$^{-1}$. \textbf{b.} Mean Sea Level seasonal range. \textbf{c.} Drifter hourly observations density in 1/4$^\circ$ bins on average for each year since 2006. Land masses are shaded white.}
\label{sfig1}
\end{figure}

Sea surface height daily data used for simulating drifter measurements are from GLORYS2V4, an ocean reanalysis product for the reference altimetry era, 1993 to present (Figure \ref{sfig1}). The GLORYS2V4 reanalysis is based on an ocean and sea-ice general circulation model (NEMO) with $1/4^\circ$ horizontal resolution, and assimilates sea surface temperature, in situ profiles of temperature and salinity, and along-track sea level anomaly observations from multiple satellite altimeters \citep{Ferry2012}. GLORYS2V4 captures climate signals and trends including GMSL rise \citep{desporte2017quality}, and realistically represents mesoscale variability \citep{garric2018performance}. These data are available from the \emph{GLOBAL-REANALYSIS-PHYS-001-031} products after registration from the Copernicus Marine Environment Monitoring Service (CMEMS) at http://marine.copernicus.eu. Sea level anomalies are calculated by subtracting the time-average of the model sea surface heights over the 1993-2015 time period.

Time-mean maps of sea surface height variance from barotropic tides and internal gravity waves at near diurnal, near semi-diurnal, and supertidal frequencies from the HYCOM numerical model were provided as electronic files by Anna Savage \citep{Savage2017}. The maps available on the original tri-pole grid of the HYCOM model were recasted in approximate 1/4$^\circ$ bins before spatial interpolation onto the drifter locations.

Significant wave height (SWH) data are from the wave model WAVEWATCH~III  from a global run at 0.5$^\circ$ spatial resolution and 3-hr temporal resolution, forced by National Centers for Environmental Prediction Climate Forecast System Reanalysis winds (NCEP CFSR), as part of the IOWAGA project [https://wwz.ifremer.fr/iowaga/, \citet{Ardhuin2011}]. These data for 1993-2015 were downloaded from the IOWAGA ftp site 

(ftp://ftp.ifremer.fr/ifremer/ww3/HINDCAST/). The SSH variance fields from surface gravity waves calculated as $\sigma^2 = \text{(SWH)}^2/16$ were subsequently interpolated linearly in time and space at the drifter locations at hourly time steps along their trajectories.

The CNES CLS15 MSS product and its formal error at 1-minute resolution is distributed by AVISO (https://www.aviso.altimetry.fr/en/data/products/auxiliary-products/mss.html). This MSS product has been derived using 20 years of data from 1993 to 2012, as an update of the CNES CLS11 product \citep{Schaeffer2012}. The formal error from the optimal interpolation mapping technique takes into account altimetric noise, ocean variability noise, and along-track biases.

The DTU15 MSS and its formal error at 1-minute resolution \citep{Andersen2015} were obtained from the ftp server of the Danish National Space Center (ftp://ftp.space.dtu.dk).

\section{Details of methods}
\label{A2}

A global or regional mean value $\overline{x}$, from individual measurements $x_i$ of a variable $x$ can be calculated as a simple arithmetic mean, or as an area-weighted mean as
\begin{equation}
\label{eq:mean}
\overline{x} = \frac{\sum_i w_i x_i}{\sum_i w_i},
\end{equation}
where $w_i = \cos \phi_i$  with $\phi_i$ the latitude of measurement $x_i$. This method intends on correcting for the convergence of satellite tracks for altimetry data, or the convergence of meridians for model data given on a regular latitude-longitude grid \citep{Masters2012,Henry2014}. Since this is the case for the GLORYS2V4 data, eq.~(\ref{eq:mean}) is used in this study to calculate the model GMSL time series. From the simulated sea level data from drifters, I have computed the GMSL in three ways: (i) using a simple arithmetic mean, (ii) using eq.~(\ref{eq:mean}) which implies that each drifter measurement is representative of a square latitude-longitude bin, and (ii) using a third method that consists in first averaging data in square bins, and second averaging all bin values using eq.~(\ref{eq:mean}). All three methods result in decadal linear trend estimates for the period 2006-2015 that are indistinguishable within error bars. As such, for consistency with the method applied to the model, I use eq.~(\ref{eq:mean}) to compute GMSL at daily time steps from simulated sea surface height data from drifters, selecting data equatorward of 66$^\circ$ and in waters deeper than 120 m.

 If $\sigma^2_i$ is the independent error variance of individual measurement $x_i$, then the variance of $\overline{x}$ defined by  eq.~(\ref{eq:mean}) is
\begin{equation}
\label{eq:var}
\text{Var}[\overline{x}] = \left(\frac{1}{\sum_i w_i}\right)^2 \sum_i w_i^2 \sigma_i^2,
\end{equation}
the square root of which provides the standard error of  $\overline{x}$ (1-$\sigma$). The variances of the drifter GMSL at daily time steps are calculated using this formula with $\sigma_i^2$ taken as the estimated individual variances from unresolved geophysical processed (surface gravity waves, tides and internal waves, and the formal CLS15 MSS error) and the instrumental error from antenna biases (Figure 1b).

Linear trend estimates and their corresponding standard error estimates for GMSL times series are calculated following the methodology of \citet{bos2008fast} and \citet{bos2013fast}, using the Hector software package available at http://segal.ubi.pt/hector/. The method estimates the linear trend in a time series with serially correlated noise assuming a model for the noise, and using a Maximum Likelihood Estimation method to estimate the trend and noise parameters. This method is preferred to the method of ordinary least squares which generates unreliable confidence intervals when the underlying assumptions that the errors are uncorrelated and with constant variance are violated, as is the case here. After I examined the power spectral densities of the residuals from linear trend estimates by ordinary least squares for each of the cases presented in the main text, I adopted the Mat{\'e}rn process \citep{Lilly2017} as a model for the noise. This model captures the spectral characteristics of the noise commonly seen for geophysical time series: a plateau at low frequencies and a roll off with approximately constant spectral slope at high frequencies.

In order to be eventually removed, the seasonal cycle of the GMSL time series are estimated by bandpass filtering. The time series are convolved with a Hanning window of 2-year length multiplied by complex exponentials at the annual and triannual frequencies. Mirror boundary conditions are applied at the end of the time series to limit the impact of edge effects. The frequencies for the seasonal filters are chosen based on the 95\% threshold of false alarm for the coherence squared between the drifter-based GMSL time series and the model GMSL time series (not shown). The smoothing filter used to extract monthly and interannual variability of the GMSL time series is a Nadaraya–Watson kernel estimator with an Epanechnikov kernel of half-width 1 month and 9 months, respectively \citep{Fan1996}. 

\begin{figure}
\centering
 \noindent\includegraphics[width=0.8\textwidth]{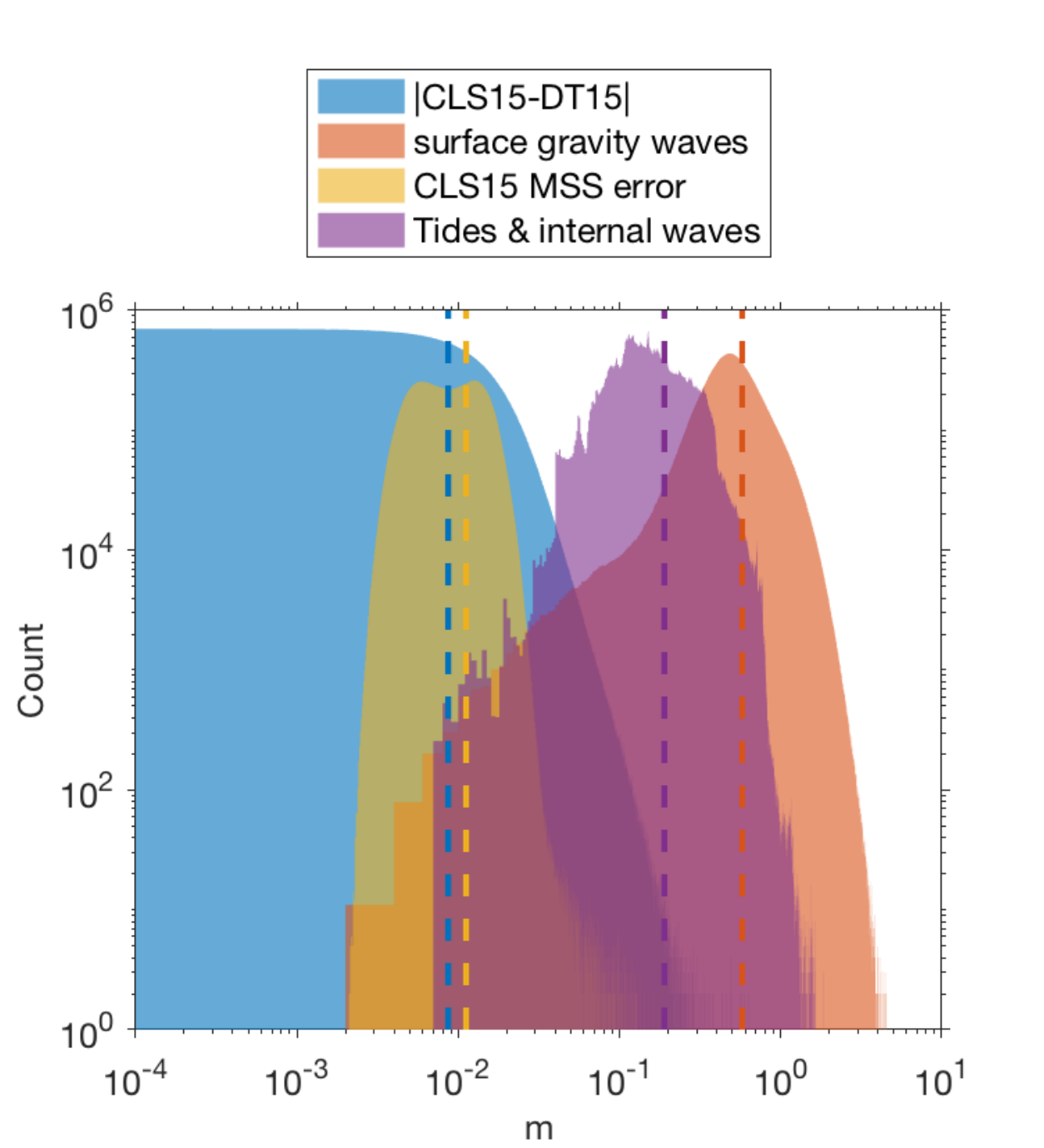}
\caption{\textbf{Distribution of error estimates}. Histograms of interpolated mean sea level errors at all drifter locations (square root of interpolated variances) from surface gravity waves, from the formal errors of the CNES CLS15 mean sea surface, and from tides and internal waves. The histogram of absolute differences between the CNES CLS15 and DTU15 mean sea surface products is also shown. Vertical dashed lines indicate the median value of each populations.}
\label{sfig2}
\end{figure}

\end{document}